\documentclass[fleqn,usenatbib]{mnras}

\usepackage[T1]{fontenc}
\usepackage{ae,aecompl}

\usepackage{graphicx}	
\usepackage{amsmath}	
\usepackage{amssymb}	
\newcommand{\Kepler}{{\it Kepler}}

\newcommand{\aCVn}{$\alpha^2$\,CVn}

\title[KIC 2569073]{Large amplitude change in spot-induced rotational modulation of the \Kepler\ Ap star KIC\,2569073}
\author[J.A. Drury et al.]{\parbox{\textwidth}{Jason A. Drury$^{1,2}$\thanks{E-mail: j.drury@physics.usyd.edu.au}, Simon J. Murphy$^{1,2}$, Aliz Derekas$^{5,6}$, \'Ad\'am S\'odor$^{5}$, Dennis Stello$^{1,2,3}$, Charles A. Kuehn$^{1,2,4}$, Timothy R. Bedding$^{1,2}$, Zs\'ofia Bogn\'ar$^{6}$, L\'aszl\'o Szigeti$^{5}$, R\'obert Szak\'ats$^{6}$, Kriszti\'an S\'arneczky$^{6}$, L\'aszl\'o Moln\'ar$^{6}$}\vspace{0.4cm}\\
$^{1}$Sydney Institute for Astronomy, School of Physics, University of Sydney, NSW, 2006, Australia\\
$^{2}$Stellar Astrophysics Center (SAC), Department of Physics and Astronomy, Aarhus University, Aarhus, DK-8000, Denmark\\
$^{3}$School of Physics, University of New South Wales, NSW, 2052, Australia\\
$^{4}$Department of Physics and Astronomy, University of Northern Colorado, Colorado, USA\\
$^{5}$ELTE Gothard Astrophysical Observatory, H-9704 Szombathely, Szent Imre herceg \'ut 112, Hungary\\
$^{6}$Konkoly Observatory, Research Centre for Astronomy and Earth Sciences, Hungarian Academy of Sciences, H-1121 Budapest,\\ Konkoly Thege Mikl\'os \'ut 15-17, Hungary\\}

\date{Accepted XXX. Received YYY; in original form ZZZ}

\pubyear{2017}

\begin{document}

\pagerange{\pageref{firstpage}--\pageref{lastpage}} 

\maketitle

\label{firstpage}

\begin{abstract}
An investigation of the 200 $\times$ 200 pixel `superstamp' images of the centres of the open clusters NGC\,6791 and NGC\,6819 allows for the identification and study of many variable stars that were not included in the \Kepler\, target list. KIC\,2569073 (V=14.22), is a particularly interesting variable Ap star that we discovered in the NGC\,6791 superstamp.  With a rotational period of 14.67\,days and 0.034-mag variability, it has one of the largest peak-to-peak variations of any known Ap star. Colour photometry reveals an anti-phase correlation between the $B$ band, and the $V$, $R$ and $I$ bands. This Ap star is a rotational variable, also known as an \aCVn\, star, and is one of only a handful of Ap stars observed by \Kepler. While no change in spot period or amplitude is observed within the 4-year \Kepler\ timeseries, the amplitude shows a large increase compared to ground-based photometry obtained two decades ago.
\end{abstract}


\section{Introduction}

Chemically peculiar A stars, in short Ap stars, are a spectroscopic subclass of A-type stars. They show markedly enhanced absorption in lines of strontium, chromium, europium, silicon and some rare-earth elements. Strong magnetic fields are also present in Ap stars \citep{Babcock1947} and often concentrate these over-abundances into spots at the magnetic poles. The presence of strong magnetic fields also results in magnetic braking \citep{Stepien2000}, reducing the rotation rate of Ap stars relative to normal A-type stars \citep{Abt95}.

Misaligned magnetic and rotational axes are a common property of Ap stars and are generally described with reference to the oblique rotator model \citep{Babcock1949, Stibbs1950}. The angle of inclination between these axes combined with the spots at the magnetic poles can result in both spectroscopic and photometric variability \citep{Wolff1983, Smith1996}. Stars with such spots that show photometric rotational modulation are classified as \aCVn\, variables. This variability can be used to determine the rotation period of the star.

\cite{Stepien2000} concluded that any angular momentum evolution must occur in Ap stars during the pre-main sequence phase (See also \citealt{Hubrig2000, North98}). Furthermore, he showed that magnetic field strength is the critical factor for angular momentum loss in Ap stars, determining the dominant spin-down mechanism. The distribution of rotation periods provides insight into the pre-main sequence evolution of Ap stars, particularly the initial angular momentum of the protostellar disk and the magnetic field strengths of these stars.

\cite{Renson2009} produced a catalogue containing $\sim$\,2000 confirmed and $\sim$\,1500 probable Ap stars over the entire sky, showing these stars are common. However, few of these stars have long-term continuous photometry that would allow for a detailed analysis of their stellar properties. 

At high photometric precision some Ap stars show rapid oscillations (roAp stars) with periods of 5-25\,minutes \citep{Kurtz1982}, but the majority are non-oscillating (noAp) stars. It should be noted that the classification of a noAp star relies only on the non-detection of oscillations. Such oscillations may be present but fall below the detection limit. Only 61 roAp stars have been identified to date \citep{Smalley15} making these relatively rare. Asteroseismology of roAp stars is important as the driving mechanism is still not fully understood and is an active area of research. 

The \Kepler\, space telescope has revolutionised the study of variable stars, advancing the detection limit for oscillations down to just a few micromagnitudes. This makes it possible to classify noAp stars more precisely than ever before. \Kepler\, observations of 7 previously known Ap stars detected pulsations in only one \citep{Balona2011a}. However, 4 new roAp stars with \Kepler\, observations have been identified; 3 from their pulsation signatures \citep{Balona2011b, Kurtz2011, Smalley15}, and 1 from a combined analysis of \Kepler\, and SuperWASP data \citep{Holdsworth2014}. 

With the exception of the roAp star from the SuperWASP project with a pulsational amplitude of 1.4\,mmag, all of the roAp stars in the \Kepler\, field exhibit pulsation amplitudes below 0.1\,mmag that would be undetectable with ground observations. Based on this, \cite{Balona2011b} proposed that noAp stars may exhibit rapid oscillations which are simply below the detection threshold.

In this paper we present an analysis of the Ap star KIC\,2569073 (V694 Lyr, \citealt{Mochejska03}). Initially identified as a variable star in $I$ and $B$ band photometry with a period of 15.26\,d (V68, \citealt{Kaza13}), KIC\,2569073 was later reported to be constant in $V$ band photometry \citep{deMarchi07}. \cite{Mochejska03} conducted a long-term variability survey of NGC\,6791 with cluster membership estimates. However, for KIC\,2569073 they did not provide a membership estimate, instead they noted its position 4 mag above and 0.3 mag blue of the turn off for cluster members, making it unlikely to be a cluster member. 

We first outline the extraction of the \Kepler\, data for KIC\,2569073 from the `superstamp' images, the characteristics of these extracted data, and the lightcurve corrections we have applied. The remainder of Sect. \ref{sec:2} describes the characteristics and calibration of a spectrum and colour photometry we have obtained. In Sect. \ref{sec:3} we discuss the classification of this star, detailing the rotation period, variation in lightcurve shape and the search for pulsation signatures in the lightcurve.

\section{Observations}
\label{sec:2}

\subsection{Kepler Observations}
\label{sec:2.1}
The \Kepler\, space telescope was primarily designed to detect transiting exoplanets and to determine the occurrence rate of small planets in the Milky Way. This required long, well-sampled, high precision time series, which are also critical for studies of intrinsic stellar variability.

\Kepler\, had a fixed observational window covering a 105 square degree field of view within the Cygnus and Lyra constellations, which was observed at a duty cycle $>$90\% between 2009 May and 2013 May. Due to limited onboard storage and connection bandwidth for data download, only specific preselected small `postage stamps' of pixels around each `target star' were selected. As a result, no data were recorded for the majority of stars within the field. In addition, \Kepler\, obtained 200 $\times$ 200 pixel `superstamp' images of the centres of the open clusters NGC\,6791 and NGC\,6819 using its long-cadence (LC) mode of 29.42\,minutes \citep{Koch2010}. We show an image of the NGC\,6791 superstamp in Figure \ref{fig:source}(a). These superstamps allow us to identify and study many variables that were not included in the \Kepler\, target list. The data are divided into quarters: Q1($\sim$34\,d), Q2-Q16($\sim$90\,d each) and Q17($\sim$33\,d), which, with the small inter-quarter gaps of up to 24\,hr, total a span of $\sim$1410\,d of observations, beginning at BJD 2454964.513.

\subsubsection{Data Extraction and Processing}
\label{sec:2.1.1}
\begin{figure}
\begin{center}\includegraphics[width=0.91\linewidth]{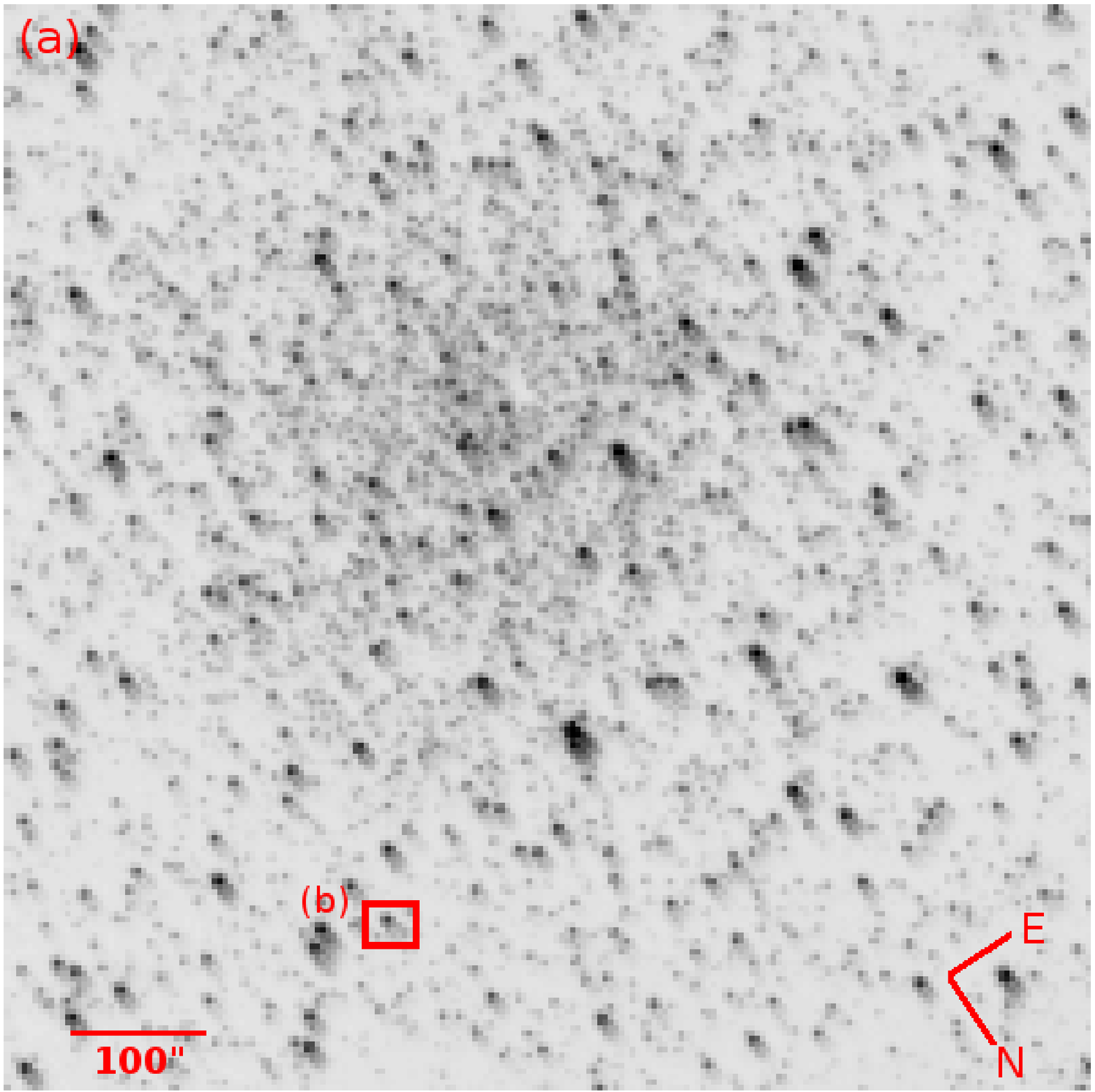}\\
\includegraphics[width=0.5\linewidth]{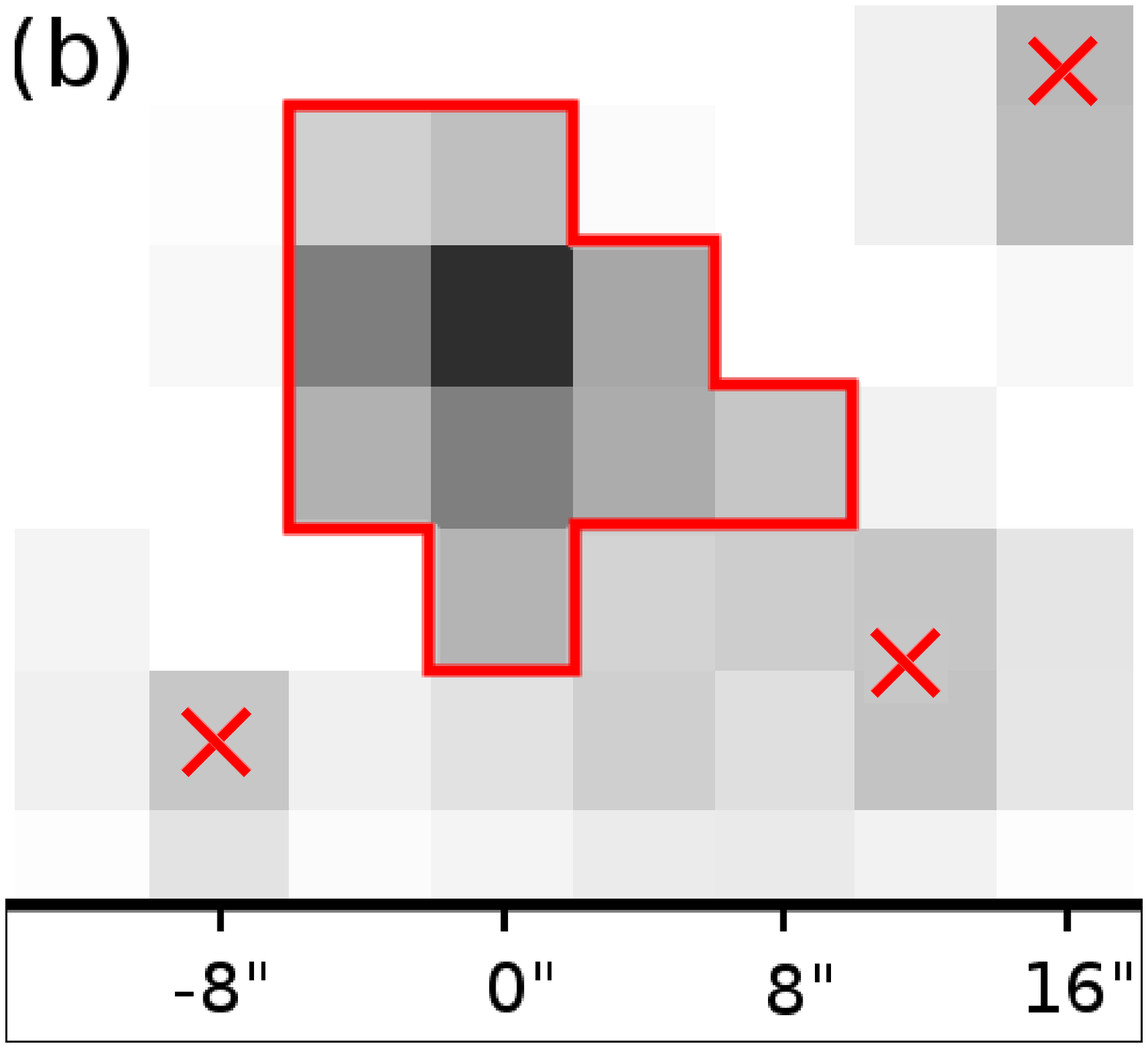}\end{center}
\caption{(a) Position of KIC\,2569073 on the superstamp of NGC\,6791. (b) Subset of pixels around target and example of the custom aperture used to extract the photometric data. Nearby stars are marked with x's}
\label{fig:source}
\end{figure}

We generated lightcurves for 25 stars with no nearby contaminants from the superstamp around NGC\,6791 using custom defined apertures for each quarter \citep{Charles2014}. To generate these custom apertures for each star, we located the brightest pixel of the star in the first image of each quarter and manually inspected all pixels in a square aperture of $\pm$ 2 pixels. We included all pixels with a fractional flux of at least 3\% of the star's brightest pixel that were not associated with a neighbouring star. To ensure a pixel was not associated with a nearby star, we produced a lightcurve and Fourier transform for every pixel within a 9$\times$9 square pixel aperture centred on the star's brightest pixel, and manually compared each Fourier transform to the Fourier transform of the brightest pixel. Any pixels with a Fourier transform containing a signal from a contaminant star greater than 3 sigma above the noise level of the target star were excluded. We summed the flux within the aperture for each image. From the lightcurves of these 25 stars we found one star with an almost-sinusoidal variability: KIC\,2569073. Figure \ref{fig:source}(b) shows its custom aperture for Q1. 

KIC\,2569073, (J2000.0, 19 20 30.799, +37 50 55.00) (Fig. \ref{fig:source}), has a $B-V$ colour of 0.54\,mag and an apparent magnitude ($V$) of 14.219\,mag \citep{Stetson2003}. Both the GAIA DR1 and SDSS databases list three nearby faint stars, 14.47$''$, 16.59$''$ and 16.93$''$ distant from KIC\,2569073 respectively. We have marked their positions in Figure \ref{fig:source}(b) with crosses. The possibility of contamination from these stars was considered and addressed through the pixel mask selection criteria above.

\begin{figure*}
\begin{center}
\includegraphics[width=\linewidth]{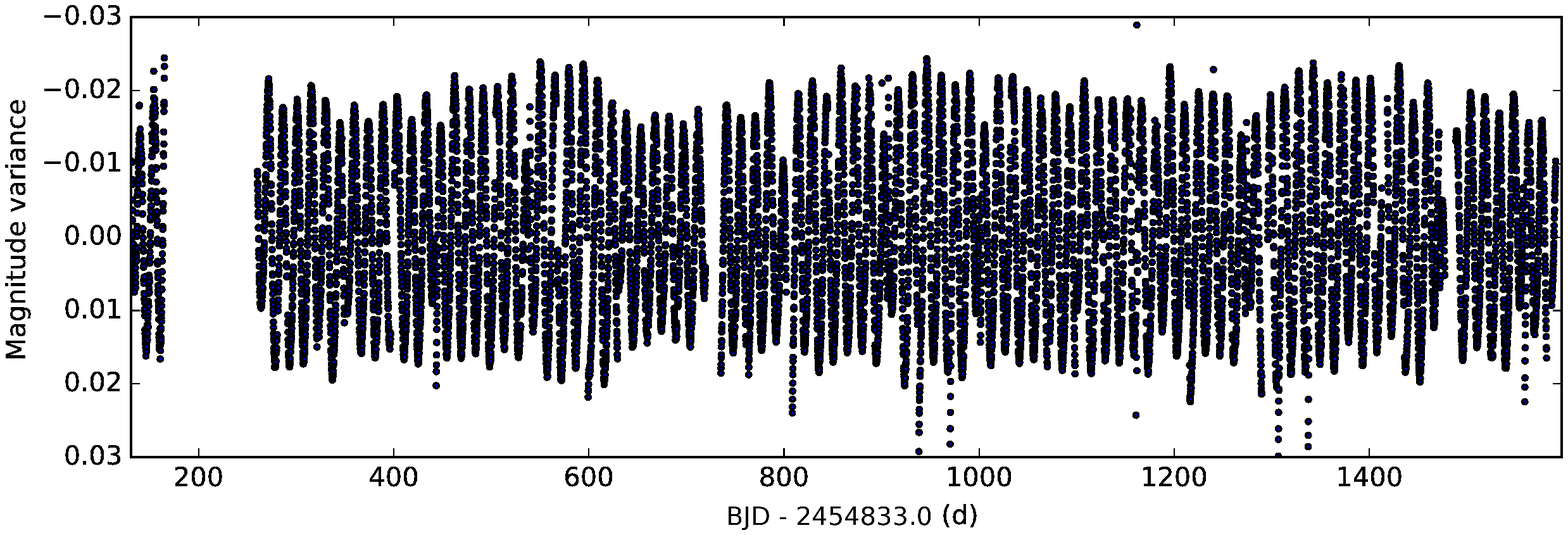}\\ 
\includegraphics[width=\linewidth]{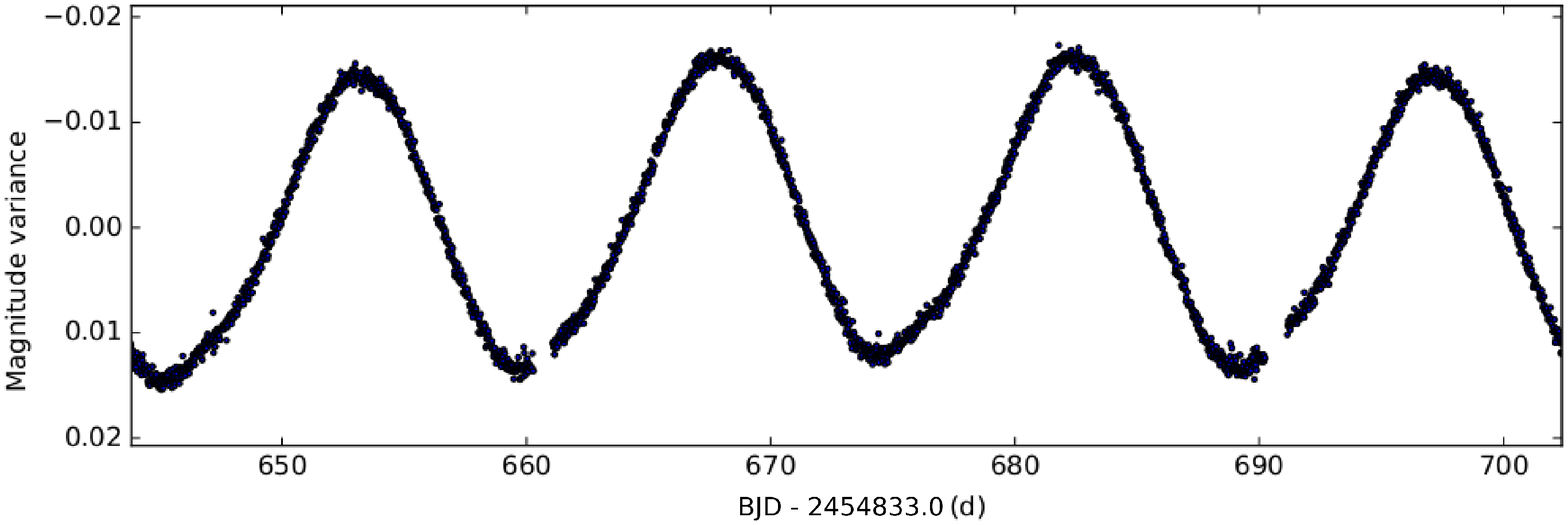}
\caption{The upper panel shows the full lightcurve of KIC\,2569073 after de-trending. In the lower panel is a zoom of the lightcurve showing its general shape.}
\label{fig:1}
\end{center}
\end{figure*}

Our raw lightcurve suffers from a number of systematic flux perturbations due to instrumental effects including the telescope's 90 day re-orientation period. To correct for these systematics, we fitted and removed fourth-order non-linear trends from each quarter. During quarter 2 (Q2), the spacecraft experienced two operational shutdown periods or safe modes. These safe modes resulted in intra-quarter trends in the data that we could not correct for easily, and we decided to discard the Q2 data set from the analysis. The remaining 16 quarters were concatenated to produce the final lightcurve. In Figure \ref{fig:1}(a) we show the full corrected lightcurve of KIC\,2569073. 

The 1st, 6th and 15th quarters show drifts in the median magnitude of the lightcurve over the length of the quarters, possibly resulting from incomplete corrections of systematic effects. Furthermore, we note the presence of correlated shifts in the amplitude of the rotational modulation of the star, but are unable to distinguish if this is intrinsic or a result of the drifts in median magnitude.

\subsection{NOT Observations}
\label{sec:2.2}

We obtained a ground-based, low-resolution (R$\sim$2000) spectrum of the star using the ALFOSC \'echelle spectrograph on the 2.5-m Nordic Optical Telescope (NOT) on La Palma on 2016 May 28. Our 20-min exposure using the \#18 grism was centred at 4360\,\AA, with a spectral range of 3450\,\AA~ to 5350\,\AA. The spectrum was reduced using the NOT standard pipeline by the on-site observer. We extracted the final spectrum by summing the flux in the central 3 pixels of the slit. The spectrum was wavelength calibrated using the Balmer series absorption lines.

\begin{figure*}
\begin{center}
\includegraphics[width=\linewidth]{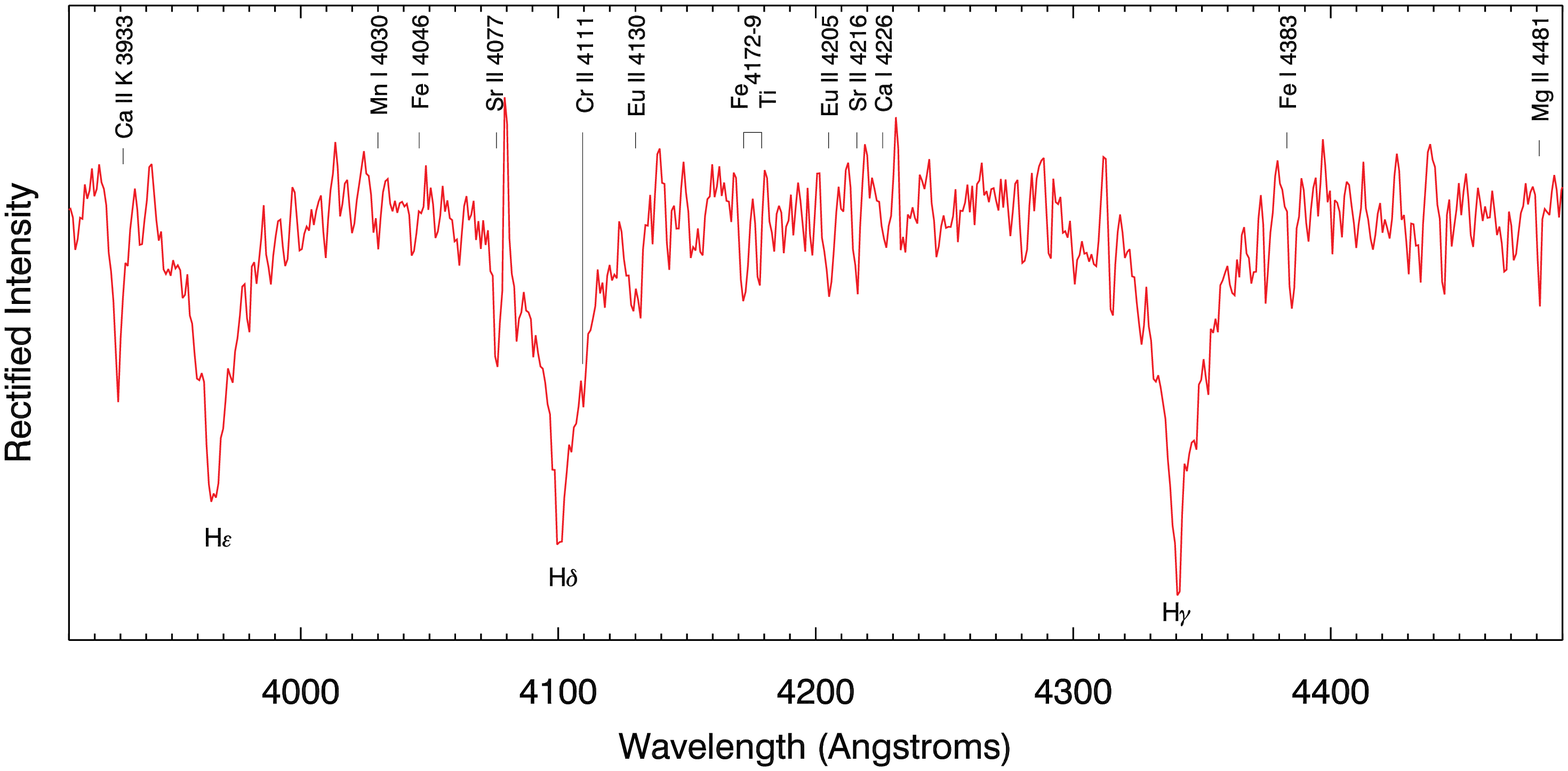}\\
\caption{The blue-violet spectrum of KIC\,2569073, showing the typical features of an A5\,Vp SrCrEu star (see \citealt{GrayaaCorbally2009}), and a broad-but-shallow Ca\,{\sc ii}\,K line. Spectral lines relevant to the spectral classification are labelled.}
\label{fig:Spectrum}
\end{center}
\end{figure*}

\subsection{Colour Photometry}
\label{sec:2.3}

We took multicolor CCD photometry with the 0.6-m Schmidt telescope at Piszk\'estet\H o Observatory, Hungary. We obtained 120 frames on 25 nights between 2015 April and 2015 July using Johnson/Bessell $B, V$ and Cousins $R_{\rm C}, I_{\rm C} $ filters. The telescope was equipped with an Apogee ALTA-U 4k$\times$4k CCD camera. On each night, images were mostly obtained in blocks of 2-3 frames per filter. The images were reduced with IRAF following the standard processing steps of bias subtraction and flat-field correction.  Aperture photometry for the target and other field stars were performed on each image using the IRAF qphot task. We used the average magnitude of seven stars as a comparison for the differential magnitude.  

\begin{figure*}
\begin{center}
\includegraphics[width=\linewidth]{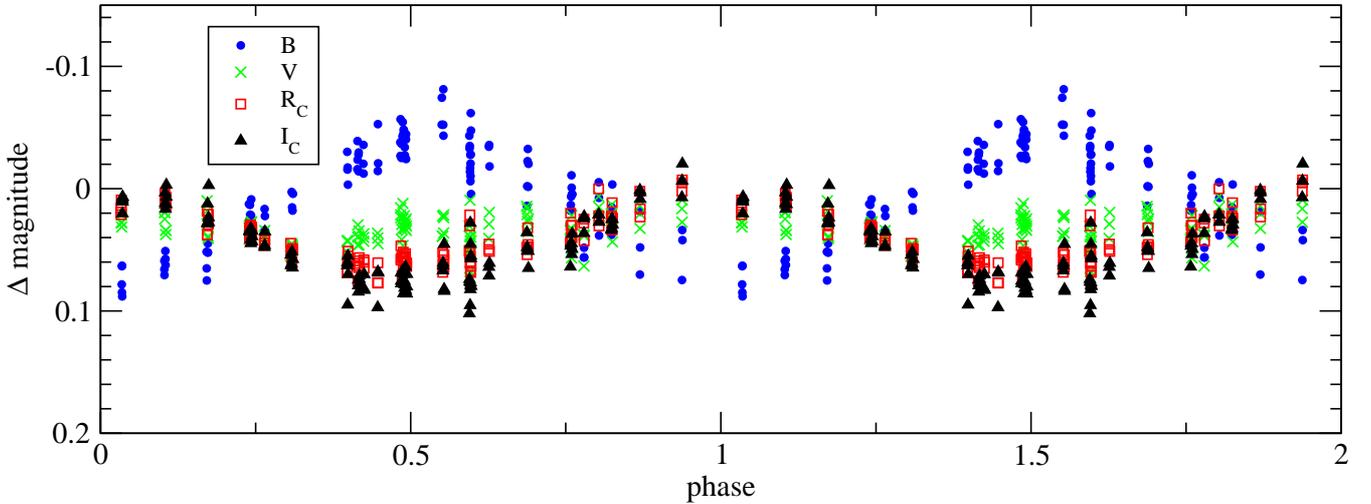}
\caption{Colour photometry of KIC\,2569073 showing anti-phase correlation between B and V, R$_C$ \& I$_C$ pass bands.}
\label{fig:BVRI}
\end{center}
\end{figure*}

\section{Results}
\label{sec:3}
\subsection{Classification and Properties}
\label{sec:3.1}

\subsubsection{Spectral Classification (A5 Vp SrCrEu)}
\label{sec:3.1.1}

We compared the spectrum with MK spectral standards, from which we determined a hydrogen line type of A5 V (Fig.\,\ref{fig:Spectrum}). We identified all of the typical peculiarities of an Ap SrCrEu star \citep{GrayaaCorbally2009}. These are evident in spite of the low signal-to-noise ratio of the spectrum, and a flux spike just redward of the Sr II 4077 line. The Ca lines are notably broad but shallow, which is often seen in magnetic Ap stars with stratified atmospheres. This spectral classification agrees with the \Kepler\, Input Catalog's  $T_{\rm eff}$  of $\sim$7420 K.

Combined with the historical colour photometry, we are able to confirm this star as a non-member of NGC\,6791 as its colour and magnitude suggest it is a foreground object.

\subsubsection{Kepler Observations}
\label{sec:3.1.2}
The lightcurve for KIC\,2569073 shows clear variability with a peak-to-peak amplitude of approximately 0.034\,mag and a primary period of $14.6679(3)$\,d (Fig. \ref{fig:1}). This signature shows at least 10 harmonics in the periodogram and is interpreted as rotational modulation, placing KIC\,2569073 as an \aCVn\, variable.

\subsubsection{Colour Photometry}
\label{sec:3.1.3}

We folded the colour photometry with the rotation period we determined from the \Kepler\, data (Figure \ref{fig:BVRI}). This revealed clear anti-phase variations between the $B$ lightcurve and the $V$, $R_C$ and $I_C$ lightcurves and shows a peak-to-peak amplitude pattern similar to the one found by \cite{Kurtz1996} for HD\,6532, namely that the $B$, $R_C$ and $I_C$ amplitudes are larger than those of the $V$ band. This phenomenon is rare within the visible wavelengths but has been well documented in \aCVn\, stars between visible and UV observations. \cite{Molnar1973} suggested this phenomenon is caused by the redistribution of flux from rare-earth line-blanketing between the $V$ and $B$ filter wavelengths. We present these peak-to-peak amplitudes along with the colour photometry obtained by \cite{Mochejska03} in Table \ref{tab:colphot}. We have converted their values from semi-amplitudes to peak-to-peak amplitudes for ease of comparison.

We note that both values show similar trends with the peak-to-peak amplitude being greater in the $B$ and infrared pass bands than the $V$ band. It should also be noted that the infrared passbands of $I$ and $I_C$ are not directly comparable. The \cite{Mochejska03} values, obtained between 1996 and 2002, are significantly smaller than those presented in this work and indicate large, long-term changes in the peak-to-peak rotational modulation amplitude and thus changes in the structure, size or location of the star spots.

\begin{table}
	\center
	\begin{tabular}{ c | c | c}
		\hline
	 	Band		&	Peak-to-peak 		&	Peak-to-peak amplitude (mag)	\\
	 			&	amplitude (mag)	&	\citep{Mochejska03}	\\
		\hline
		$B$		&	0.13			&	0.066			\\
		$V$		&	0.03			&	0.006			\\
		$R_C$		&	0.07			&	--			\\
		$I_C$		&	0.28			&	--			\\
		$I$		&	--			&	0.062			\\
		$K_p$		&	0.34			&	--			\\
		\hline
	\end{tabular}
	\caption{Comparision of peak-to-peak amplitudes of the $B$, $V$, $R_C$ and $I_C$ photometric observations.}
	\label{tab:colphot}
\end{table}

Importantly, the $V$ amplitude of KIC\,2569073 matches the amplitude observed by \Kepler\ as expected due to the similarities in the passbands. Meanwhile, the increase in rotational modulation amplitude from $V$ to $R_C$ to $I_C$ band is now observed for magnetic Ap stars across a range of spectral types from late-B \citep{Grobel2017} through early-A \citep{Kurtz1996} to mid-A stars (this work). 

\subsection{Pulsation signatures}
\label{sec:3.2}

The spectral class and temperature of KIC\,2569073 are similar to those of known roAp stars \citep{Smalley15}. RoAp stars typically have pulsation periods between 5 and 25 minutes corresponding to pulsation frequencies above the \Kepler\, LC Nyquist frequency. \cite{Murphy12} noted that the $\sim$30 minute integration time for the LC \Kepler\, data results in amplitude reduction of pulsation frequencies beyond the Nyquist limit, but these pulsations can still be investigated \citep{Murphy13}. The amplitude reduction is described by equation \ref{eqtn:ampl} where $A_0$ is the true amplitude, the observed amplitude is $A$ and n is the number of observations per oscillation cycle. 

\begin{equation}\label{eqtn:ampl}
	A = \cfrac{\sin(\pi/n)}{\pi/n}A_0
\end{equation}

\begin{figure}
\begin{center}
	\includegraphics[width=\linewidth]{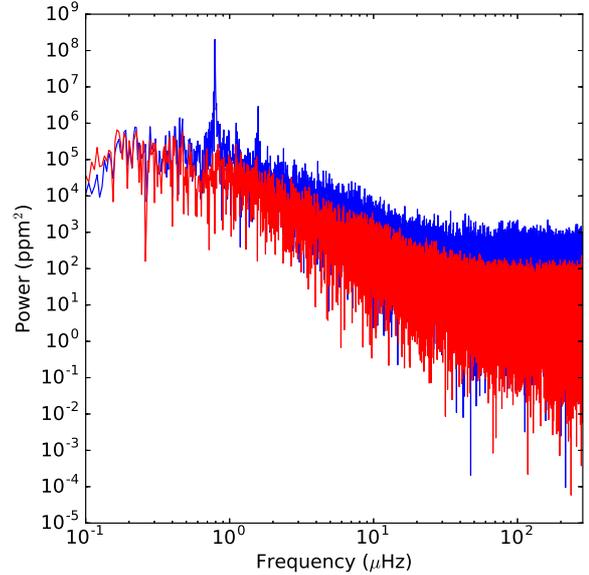}
\caption{Logarithmic power Spectrum of KIC\,2569073 (blue), overlaid with the spectrum after removing the main oscillation period and its first fifteen harmonics (red).}
\label{fig:4} 
\end{center}
\end{figure}

To search for pulsation signatures, we identified the rotational frequency as $\mathrm{\nu_{rot} = 0.78909104\,\mu}$Hz using \texttt{PERIOD04} \citep{Lenz05} and fitted and subtracted the first 15 harmonics of this frequency (Table \ref{tab:freqs}). The Fourier transform of the residuals is displayed in Fig. \ref{fig:4}. We were unable to identify any significant pulsation signatures above the noise level of 15\,$\mu$mag up to a frequency of 3500\,$\mu$Hz. It is possible however that there are pulsations in this star below this level. We calculated upper limits for the true amplitudes of any rapid oscillations to be 750\,$\mu$mag and 90\,$\mu$mag for the 5 and 25-min pulsation periods respectively. These limits are much higher than the noise level due to amplitude attenuation caused by undersampling (Eq. \ref{eqtn:ampl}).

\begin{table}
	\center
	\begin{tabular}{ c | c | c | c | c | c }
		\hline
		\hline
	 	ID	 	&      Frequency 		 &	Amplitude ($\sigma$) \\
	 			&      $\mu$Hz	 	 &	mag			 \\
		\hline
		f$_1$		&	0.78909104		 & 0.01718 	 \\
		2f$_1$		&	1.57818208		 & 0.00149 	 \\
		3f$_1$		&	2.36727313		 & 0.00039 	 \\
		4f$_1$		&	3.15636417		 & 0.00017 	 \\
		5f$_1$		&	3.94545521		 & 0.00011 	 \\
		6f$_1$		&	4.73454625		 & 0.00002 	 \\
		7f$_1$		&	5.52363729		 & 0.00004 	 \\
		8f$_1$		&	6.31272833		 & 0.00004 	 \\
		9f$_1$		&	7.10181937		 & 0.00005 	 \\
		10f$_1$		&	7.89091042		 & 0.00004 	 \\
		11f$_1$		&	8.67995701		 & 0.00006 	 \\
		12f$_1$		&	9.46904402		 & 0.00002 	 \\
		13f$_1$		&	10.25813102		 & 0.00004 	 \\
		14f$_1$		&	11.04721801		 & 0.00003 	 \\
		15f$_1$		&	11.83630501		 & 0.00002 	 \\
		\hline
	\end{tabular}
	\caption{Frequencies and amplitudes of the identified rotational frequency peaks in the frequency spectrum of KIC\,2569073. All amplitudes are determined to $\pm$ 0.00002\,mag.}
	\label{tab:freqs}
\end{table}

To ensure we are able to detect stellar pulsations in \Kepler\, LC data for roAp stars we downloaded the MAST LC data files of the four known roAp stars (KIC\,10483436, KIC\,10195926, KIC\,4768731, KIC\,7582608) which have pulsation modes identified in the SC data. We subjected these to the same systematics correction procedure as KIC\,2569073. Once again we used \texttt{PERIOD04} to search for pulsation signatures up to 3500\,$\mu$Hz. Only two of these known roAp stars, KIC 10195926 and KIC 7582608, show pulsation signatures above the noise level in the LC data, with the remaining two stars' amplitudes being too heavily attenuated by undersampling in LC for detection. These stars also have the highest oscillation amplitudes in the SC data. This suggests noAp stars may indeed have rapid oscillations which are simply below the limit of detection in LC data. Therefore, we cannot definitively rule out rapid oscillations in KIC\,2569073, with amplitudes below our quoted limits. We have included the four roAp stars and their frequencies and amplitudes in Table \ref{tab:roAp}.

\begin{table}
	\center
	\begin{tabular}{ c | c | c}
		\hline
	 	KIC ID		&	Frequency ($\mu$Hz) 	&	Amplitude (mmag)	\\
		\hline
		10483436	&	1353			&	0.068			\\
		10195926	&	972.6			&	0.176			\\
		4768731		&	711.2			&	0.062			\\
		7582608		&	2103.4			&	1.45			\\
		\hline
	\end{tabular}
	\caption{Rapid oscillation frequncies of known \Kepler\ roAp stars and their amplitudes from SC data.}
	\label{tab:roAp}
\end{table}

\subsection{Rotation Period Variations}
\label{sec:3.3}

The rotation of some Ap stars appears more complicated with evidence of rotational period variations occurring over timescales of decades \citep{Krticka17}. Whilst these variations were of periodic nature and explained by torsional oscillations within the star, no short-timescale studies have been conducted to investigate the stability of these rotation periods. With the unique quality and quantity of data of an \aCVn\, star from \Kepler, we conducted an observed-minus-calculated (O-C) analysis of the lightcurve. This constitutes the first in-depth analysis of the stability of the rotation period of an \aCVn\, star and is important in searching for possible spot variation on short timescales.

We determined the phase variations by a template-fitting O$-$C method \citep{Sodar2017} that takes into account the shape of the full rotationally modulated lightcurve.

\begin{figure}
\begin{center}
\includegraphics[clip, trim=0cm 0cm 1.7cm 0.2cm ,width=0.90\linewidth]{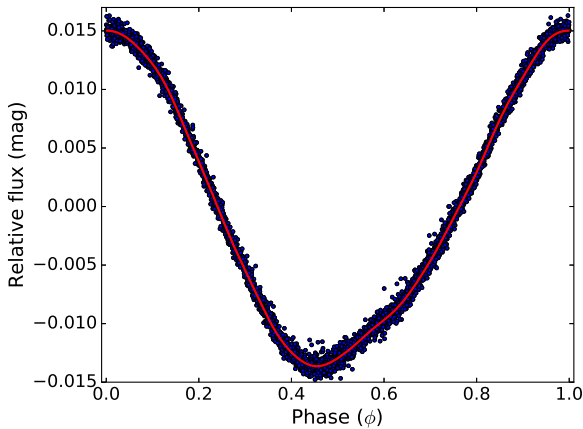}
\caption{Template rotation curve for KIC\,2569073 (red) used to calculate the O$-$C diagram with 90\,d segment of lightcurve starting at 630\,d and phase-folded on the rotation period (blue).}
\label{fig:template}
\end{center}
\end{figure}

To describe the shape of the rotationally modulated lightcurve unaffected by amplitude, phase and zero-point variations, we fitted 15 harmonics of the rotation frequency (0.0681757\,d$^{-1}$) to a 90\,d section of the data starting at 630\,d (approximately the section shown in Fig. \ref{fig:1} (b)), where lightcurve-shape variations are negligible. We refer to the result of this Fourier-fit as the template rotation curve (Fig. \ref{fig:template}).

For the calculated times of maxima (C), we used the ephemeris $$ T_\mathrm{max} = T_0 + 14.6680\,\mathrm{d}\cdot E,$$ \noindent where $E$ is the epoch number, showing the elapsed rotation cycles since the reference epoch, $T_0 = 257.00\,\mathrm{d}$ ($\mathrm{BJD}\,\, 2455090.00$). The reference epoch was selected to avoid the presence of large data gaps in the lightcurve.

\begin{figure}
\begin{center}
\includegraphics[clip, trim=0cm 0cm 1.7cm 0.2cm ,width=0.9\linewidth]{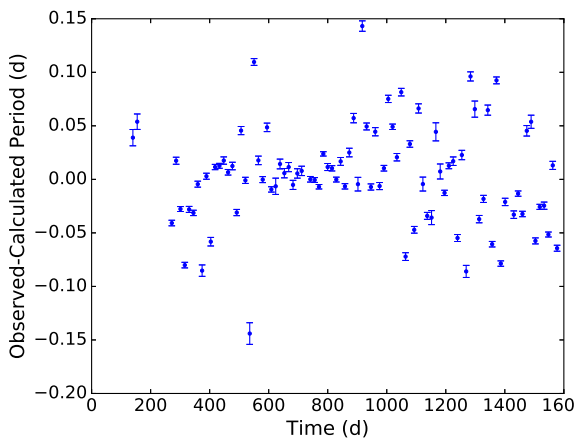}
\caption{O$-$C diagram of KIC\,2569073 based on the template lightcurve.}
\label{fig:3}
\end{center}
\end{figure}

The observed times of maxima (O) were determined by fitting the phase shift, along with an amplitude scaling factor and a magnitude zero point, of the template rotation curve to one rotation-period-long lightcurve segments. We omitted segments with poor rotation-phase coverage caused by gaps in the data. The obtained O$-$C diagram (Fig. \ref{fig:3}) contains 65 data points.

Systematic trends in the data typically have similar time scales to the rotation period. This makes decoupling intrinsic spot (lightcurve shape) variations on time scales on the order of the rotation period difficult. Our O$-$C analysis shows no coherent variations in the rotation period on time scales longer than this period.

For spot variations on shorter timescales, the lightcurve shape should change visibly from one rotation period to the next. To test this, we plotted the phase-shifted lightcurve, overlaying the rotational modulation signal upon itself. We detected some minor changes in the lightcurve shape in the vicinity of maxima and minima (e.g. at $\sim$373\,d), however the strongest period variations in these regions appear to be associated with preceeding gaps in the lightcurve and thus we have concluded these changes originate from instrumental effects. This is supported by the stability in both the lightcurve shape and the O$-$C diagram between 630\,d and 720\,d, where instrumental trends appear to be almost negligible.

We note that the standard errors of the above described fitting process take into account only short time-scale uncorrelated (assumed Gaussian) noise, but do not reflect longer time-scale instrumental variations in the data. To obtain more realistic uncertainty estimations, we repeated the fitting process with lightcurves de-trended using different order non-linear trends fit to each quarter separately. The median O$-$C deviation between the different de-trended lightcurves was calculated to be $\sim$0.11\,d. This is twice the standard deviation of the best O$-$C analysis. As such we conclude that there is no evidence for short-term period variation in the spots of KIC\,2569073. They appear to be well-anchored to the same position on the star.

\section{Summary}
\label{sec:4}
We have determined KIC\,2569073 to be an \aCVn\, star with a rotational period of 14.668\,d. Its peak-to-peak amplitude of 0.034\,mag makes it one of the most variable Ap stars in the \Kepler\ field. We notice a large change in amplitude of the rotational modulation between the Kepler data set and the one obtained by \cite{Mochejska03} between 1996 and 2002. Due to the difficulties in obtaining a Kepler light curve free from artefacts of the reduction, we were unable to say whether the pulsational amplitude remains constant over the 4-yr \Kepler\, data set. However, it is approximately constant, and changes of the amplitude seen between the Kepler data and the photometry of \cite{Mochejska03} are not replicated within the \Kepler\ data set. Further, we looked for period variations in the 4-yr \Kepler\ dataset by varying the detrending parameters in the light curve reduction and conducting O$-$C analyses. We determined the period to be constant, within the global uncertainties. These results are important in framing the time-scales of spot evolution for Ap stars.

We have also presented colour photometry, which shows an anti-phase relationship between the B lightcurve and the V, R and I lightcurves. This relationship, along with the long timescale amplitude modulation in the \Kepler\ ($\sim$ $V$) passband, suggests that KIC\,2569073 may be particularly useful for studying the formation and evolution of stellar magnetic fields and atmospheres.

\section*{Acknowledgments}
We thank the referee B. Smalley for his comments, which greatly improved this paper.

Funding for the Stellar Astrophysics Centre is provided by The Danish National Research Foundation (Grant agreement no.: DNRF106).

IRAF is distributed by the National Optical Astronomy Observatories, which are operated by the Association of Universities for Research in Astronomy, Inc., under cooperative agreement with the National Science Foundation.

This project has been supported by the Hungarian NKFI Grants K-113117, K-115709, K-119517 and PD-116175 of the Hungarian National Research, Development and Innovation Office. AD has been supported by the Postdoctoral Fellowship Programme of the Hungarian Academy of Sciences and by the Tempus K\"ozalap\'itv\'any and the M\'AE\"O. LM and \'AS have been supported by the J\'anos Bolyai Research Scholarship of the Hungarian Academy of Sciences. AD and LSz would like to thank the City of Szombathely for support under Agreement No. 67.177-21/2016.




\bibliographystyle{mnras}
\bibliography{KIC2569073} 


\bsp

\label{lastpage}

\end{document}